\documentstyle[prb,epsfig,aps,multicol]{revtex}

\begin{document}
\title{Ratchet-like dynamics of fluxons in annular Josephson
junctions \\ driven by bi-harmonic microwave fields}
\author{A. V. Ustinov, C. Coqui and A. Kemp}
\address{Physikalisches Institut III, Universit\"at
Erlangen-N\"urnberg, D-91058 Erlangen, Germany}
\author{Y. Zolotaryuk}
\address{Bogolyubov Institute for Theoretical Physics,
National Academy of Sciences of Ukraine, Kiev 03143, Ukraine}
\author{M. Salerno}
\address{Dipartamento di Fisica ``E. R. Caianiello'' and
Istituto Nazionale di Fisica della Materia (INFM), \\ Universita?
di Salerno, I-84081 Baronissi, Salerno, Italy}

\date{\today}
\maketitle

\begin{abstract}
Experimental observation of the unidirectional motion of a
topological soliton driven by a bi-harmonic ac force of zero mean
is reported. The observation is made by measuring the
current-voltage characteristics for a fluxon trapped in an annular
Josephson junction that was placed into a microwave field. The
measured dependence of the fluxon mean velocity (rectified
voltage) at zero dc bias versus the phase shift between the first
and second harmonic of the driving force is in qualitative
agreement with theoretical expectations.
\end{abstract}



\newpage

The role played by solitons in various physical systems is
important and commonly known. Experimental observation of
dynamical effects produced by solitons is  in many cases difficult 
because
real systems are very often far from the idealized mathematical
models which give rise to soliton solutions. Solitons of the most
common type are topological {\it kinks}, a well-known example
being a magnetic flux quantum (fluxon) in long Josephson junctions
(LJJs) \cite{Barone,Ust-rev98}. A fluxon in LJJ can be easily
driven by the bias current applied to the junction. The motion of a
fluxon gives rise to the dc voltage $V$ across the junction, which
is proportional to the fluxon's mean velocity. Varying the dc bias
current $I$, one can produce a dependence $V(I)$, which is the
main dynamical characteristic of the LJJ. An experimentally
obtained $V(I)$ curve allows to identify many interesting features
in the dynamics of fluxons trapped in a LJJ \cite{Ust-rev98}.

A microwave field applied to a LJJ gives rise to an ac drive
acting on the fluxon. In a spatially homogeneous lossy system, an
ac drive may only induce an oscillatory motion of a kink, which is
hard to observe in LJJ due to the absence of dc voltage. It was
first predicted \cite{Bob1} and then observed experimentally
\cite{UMal} that an ac drive can support motion of a kink with a
{\em nonzero} average velocity $\langle v \rangle$ in a system
with a periodic spatial modulation. Thus, if an inhomogeneous
spatial potential is present, the fluxon motion can be locked to
the external ac drive frequency. In this case the vertical steps
(``Shapiro steps") appear in the dependence of the average fluxon
velocity $\langle v \rangle$ on the dc bias. The fluxon velocity
in this frequency-locked state can be either positive or negative,
depending on the initial conditions. Another interesting effect
for a homogeneous system, in which the ac-driven fluxon drift may
occur in either direction but is not frequency locked, was
recently reported by Goldobin et al. \cite{Gol-Mal-Us-2002}.

Transport phenomena in nonlinear systems can also be induced by
the so called {\it ratchet effect}. This phenomenon has been
studied in a large variety of physical systems \cite{ratchet} and
manifests itself by an unidirectional motion under the influence
of forces with zero mean. The underlying mechanism is the breaking
of  those spatial and/or temporal symmetries \cite{symm} of the
system,  which connect trajectories with specular velocities (as well
as the possibility of phase-locking of the particle dynamics by
the external field \cite{Barbi-Saler-2001}). So far, great
attention has been devoted to the study of the soliton motion in
ratchet potentials. Josephson fluxon ratchets have been proposed
and designed using either continuous long junctions of special
shapes \cite{Edik-PRE01,Carapella-PRL01} or discrete junction
arrays \cite{Falo-EPL99,Trias-PRE00}, including the quantum case
\cite{Mooij-ratchet-PRL03}. In these structures, an ac drive applied
to a fluxon placed into an asymmetric (ratchet-like) spatial
potential leads to the rectified voltage at zero bias. The mechanism
underlying soliton ratchets in asymmetric potentials and driven by
harmonic forces was investigated for the double-sine-Gordon system
which can be mapped to an array of 3-junction SQUIDs
\cite{Saler-Qint-2002}. The possibility of a ratchet-like effect
induced by more complex ac drives but \emph{in the absence of any
spatial ratchet potential} was also demonstrated
\cite{Fl-Zol-Fist-2002,Saler-Zol-2002}.

The importance of the ratchet-like effect induced by a bi-harmonic
force in the context of Josephson junctions resides in the fact
that it gives a non-zero voltage state in absence of dc bias. The
resulting current-voltage characteristics of a uniform LJJ is
fully controlled by the phase difference between the two microwave
field components driving the junction, which is a nice feature for
practical applications. In a more general context, the
experimental observation of soliton ratchets induced by asymmetric
forces, first reported in this paper, opens new perspective for
soliton transport in other physical systems. In contrast to
soliton ratchets induced by asymmetric potentials, which are
generally more difficult to implement and to control (a change of
the potential requires a structural change of the system), soliton
ratchets induced by asymmetric external fields can be implemented
in any physical system described by soliton equations.

The aim of this Letter is to present the first experimental
observation of the \emph{rectified dc voltage} induced by the
unidirectional motion of a fluxon driven by \emph{bi-harmonic
microwaves} of zero average, applied to a \emph{spatially uniform}
long Josephson junction. Our results qualitatively confirm the
theoretical predictions \cite{Fl-Zol-Fist-2002,Saler-Zol-2002}.

\begin{figure}[h]
\centerline{\epsfig{file=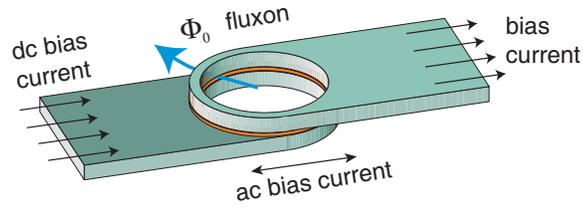,width=3.0in}} \caption{Schematic
view of an annular Josephson junction with a single fluxon trapped
in it. The junction is biased by a dc current and an ac bias,
induced by applied microwaves.} \label{scheme}
\end{figure}

Experiments have been performed using Nb/Al-AlO$_{x}$/Nb Josephson
junctions \cite{Hypres}. We have measured two samples which have
shown similar behavior; here we present data for one of them
having the mean diameter $2R=100\,\mu$m and the annulus width
$W=4\,\mu$m. The dc bias current was applied to the ring-shaped
junction via the leads having the so-called Lyngby geometry
\cite{Lyngby}(schematically shown in Fig.~\ref{scheme}). In this
configuration the bias electrode width equals the ring diameter,
and the bias current is distributed very uniformly around the
sample. This is due to the fact that the current distribution
carried by a superconducting strip is larger on the edges and its
profile $(R^2-y^2)^{-1/2}$ (where $y$ is the transverse coordinate
\cite{Ann-distrib}) exactly matches the uniform biasing of the
junction.

The critical current $I_{{\rm c}}$ of this junction in the state
with no trapped fluxons was found to be about $0.9$ of the maximum
expected value, corresponding to a nearly uniform current flow
over the junction area. The junction has the critical current
density of about $1\,$kA/cm$^2$, which corresponds to the
Josephson length of $\lambda_{\rm J}\approx 12\,\mu$m and plasma
frequency of about $f_p=120\,$GHz. This implies the normalized
junction length $2\pi R/\lambda _{\rm J}\equiv\ell \approx 26$ and
width $W/\lambda _{\rm J}<1$, i.e. the junction can be regarded as
long and quasi-one-dimensional. The measurements were done at the
temperature $4.2\,$K, using a shielded low-noise measurement
setup. The radio-frequency (rf) current was supplied by means of
an open-ended coaxial cable antenna placed above the junction. The
antenna was oriented parallel to the dc bias current leads,
therefore the inductively coupled rf current was flowing in
parallel to dc current. The applied ac power was supplied by the
two microwave sources (HP8672A and HP83620A, combined via a
directional coupler) having the frequencies $f_1=f=\omega /(2\pi)$
and $f_2=2f$ with $f$ in the range between $1$ and $5$ GHz. The
sources were phase locked to a common 10 MHz reference signal. The
phase shift $\theta$ between the sources was controlled
electronically via the phase of the reference signal and monitored
using a network analyzer. The ac power levels $P_1$ and $P_2$,
mentioned below, pertain to the input at the top of the cryostat.

Trapping of a fluxon in the junction was achieved by cooling the
sample below the critical temperature ${T_{c}} \approx 9.2\,$K for
the transition of Nb into the superconductive state, with a small
dc bias current applied to the junction \cite{UstMal}. The
residual fluxon depinning current $I_{{\rm dep}}$ was found to be
rather small, less than $3$\% of the Josephson critical current
$I_{{\rm c}}$, measured without the trapped fluxon. As a fluxon can
only be trapped by junction's local inhomogeneities in the absence
of the magnetic field, this indicates a fairly high uniformity of
the junction.

In our experimental configuration, the dynamics of the
superconducting phase difference across the junction is described
by the perturbed sine-Gordon equation
\begin{eqnarray}
\nonumber
&&\varphi_{tt}-\varphi_{xx} + \sin \varphi + \alpha \varphi _{t}
=\gamma +\tilde \gamma(t)~;~\\
&&\tilde \gamma(t)=\tilde \gamma_1\sin(\Omega t)+\tilde
\gamma_2\sin(2\Omega t+\theta), \label{sG1}
\end{eqnarray}
\noindent with the boundary conditions
\begin{equation}
\varphi(\ell)=\varphi(0)+2\pi\,;\;\;\;
\varphi_x(\ell)=\varphi_x(0)\,, \label{bc}
\end{equation}
where $x$ and $t$ are the length along the junction and the time,
measured, respectively, in units of the Josephson length
$\lambda_{\rm J}$ and the inverse plasma frequency $(2\pi
f_p)^{-1}$. The dissipation constant due to the quasiparticle
tunneling current, $\alpha$, is of the order of 0.05 in our
experiment. Here $\gamma$, $\tilde \gamma_1$ and $\tilde \gamma_2$
are the dc and ac bias current densities, both normalized to the
junction's critical current density and $\Omega=\omega/(2\pi
f_p)=f/f_p$ is the dimensionless frequency of the ac bias. The
amplitudes $\tilde \gamma_1\sim \sqrt{P_1}$ and $\tilde
\gamma_2\sim \sqrt{P_2}$ correspond to the first and second
microwave harmonics, respectively, and $\theta$ is the phase shift
between them that we controlled in this experiment.

\begin{figure}[htbp]
\centerline{\includegraphics[width=7.50cm,height=5.60cm,clip]{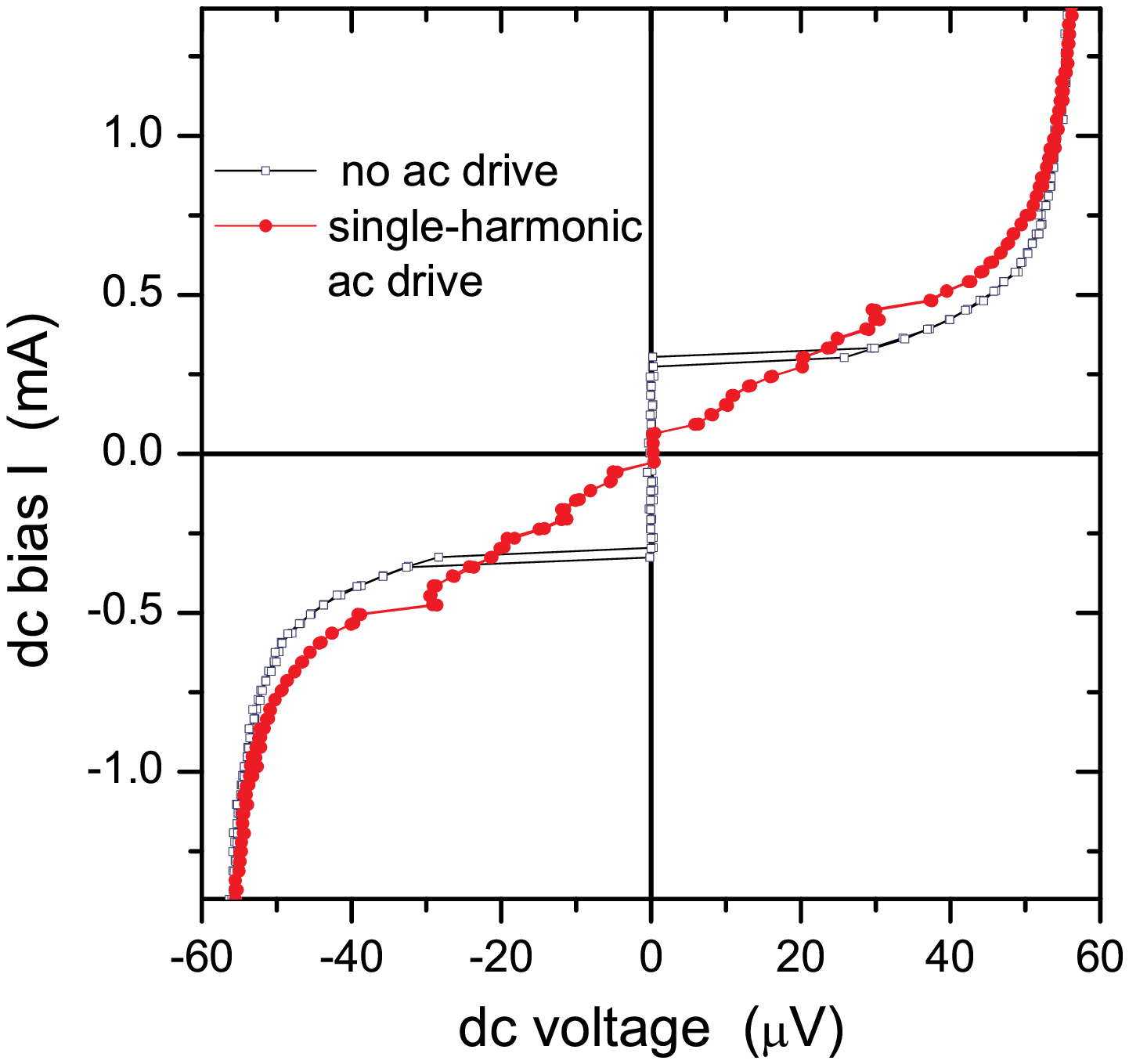}}
\centerline{\includegraphics[width=7.50cm,height=5.60cm,clip]{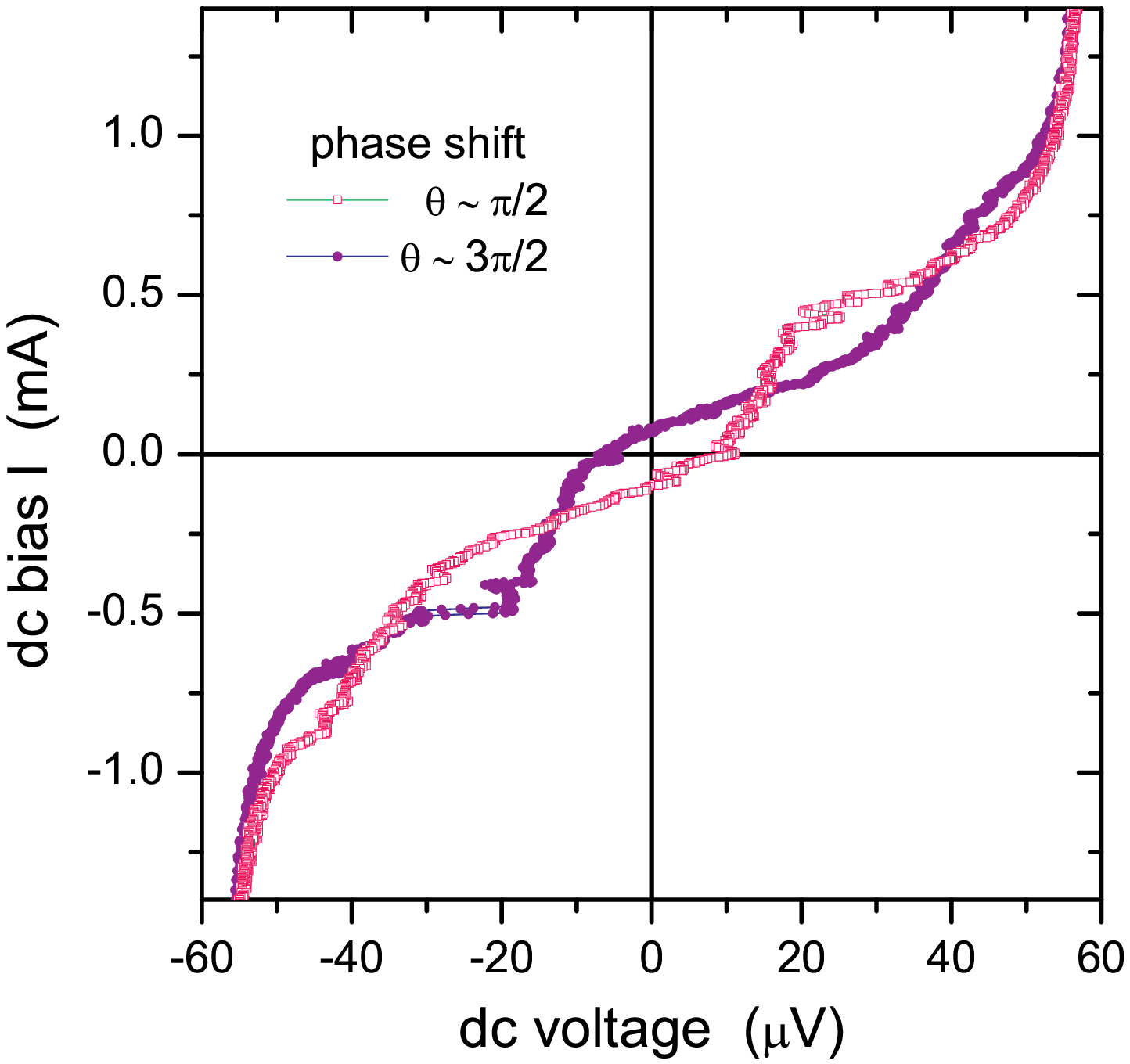}}
\caption{(a) Current-voltage characteristics of a single fluxon in
annular Josephson junction with no ac drive (open symbols) and
with single-harmonic ac drive (solid symbols) having the frequency
$f_1=4.8\,$GHz. (b) A single-fluxon current-voltage
characteristics. The fluxon is driven by bi-harmonic ac drive
having the frequencies $f_1=1.2\,$GHz and $f_2=2.4\,$GHz and power
$P_2=+4\,$dB. The phase shift $\theta$ between the two harmonics
is fixed to $\pi/2$ (open symbols) and $3\pi/2$ (solid symbols). }
\label{IVC1}
\end{figure}

The effect of a single harmonic microwave (here $\tilde
\gamma_2=0$) on the motion of a fluxon in the junction is
presented in Fig.~\ref{IVC1}a. It shows the current-voltage
characteristics of the fluxon with no ac drive (open symbols) and
with the single-harmonic ac drive (solid symbols) having the
frequency $f=4.8\,$GHz. We note that the fluxon characteristics in
the presence of a monochromatic ac drive of any frequency remains
symmetric with respect to the origin, as was observed in the
earlier experiments \cite{UMal,Gol-Mal-Us-2002}.

\begin{figure}[tbp]
\centerline{\includegraphics[width=8.cm,height=5.80cm,clip]{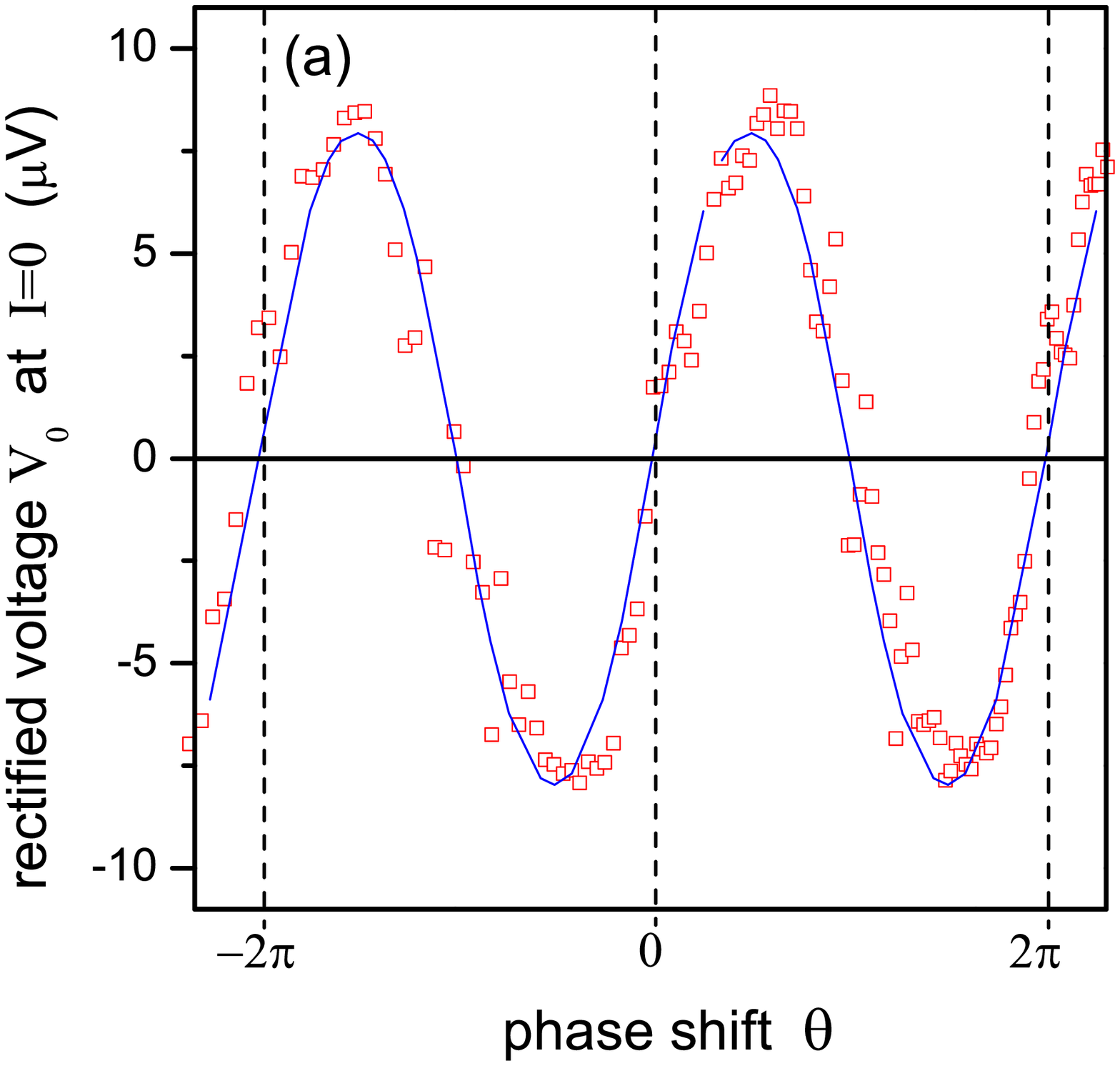}}
\centerline{\includegraphics[width=8.cm,height=5.80cm,clip]{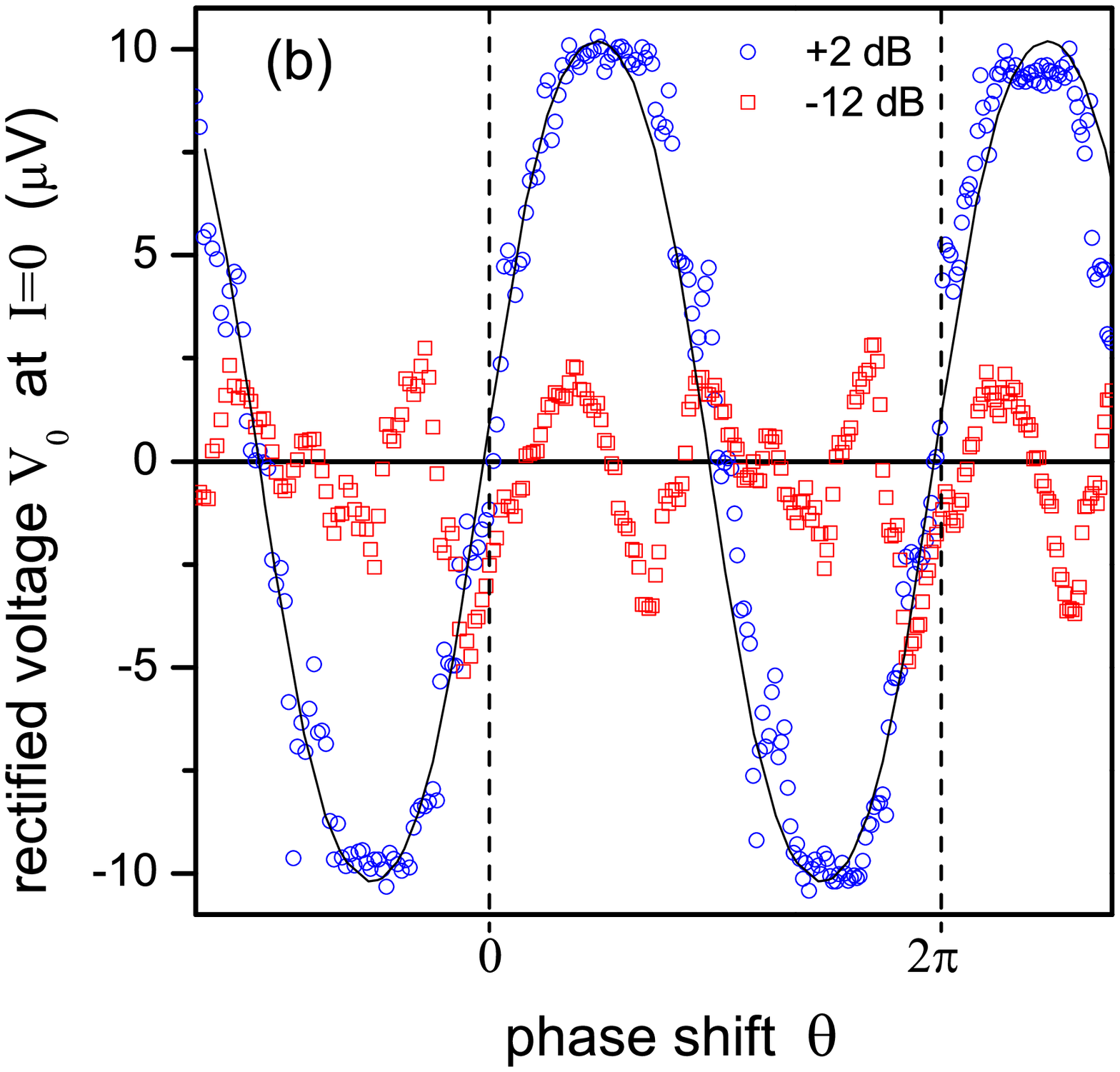}}
\caption{A dependence of the rectified dc voltage measured at zero
dc bias current on the phase shift $\theta$ between the two
harmonics of ac drive. (a) $f_1=1.2\,$GHz and $f_2=2.4\,$GHz and
power $P_2=+4\,$dB. (b) $f_1=2.4\,$GHz and $f_2=4.8\,$GHz, for two
power levels. Solid lines show results of numerical computation of
the voltage for the dimensionless ac bias amplitudes $E_1=E_2=0.1$
(a) and $E_1=E_2=0.135$ (b). } \label{phase-shift}
\end{figure}

The effect of the second harmonic in the microwave spectrum is
reported in Fig.~\ref{IVC1}b. The two curves, shown in that
figure, correspond to two different phase shifts between the
drives $P_1$ and $P_2$. The salient feature of these curves is the
non-zero voltage at zero dc current through the junction. The sign
of this rectified voltage gets reversed by changing the phase
shift $\theta$ from $\pi/2$ to $3\pi/2$ (or to $-\pi/2$). The
dependence of the rectified voltage at zero bias on the phase
shift is presented in Fig.~\ref{phase-shift}(a). We observe that
the rectified voltage reverses its sign close to $\theta=0$ and
$\theta=\pi$ and changes periodically with the phase shift. We
also measured the dependence of this effect on power and found
that, at somewhat lower power levels, the dependence of the
rectified voltage on the phase shift may have several maxima
within period of $2\pi$, see Fig.~\ref{phase-shift}(b). Possibly,
at these values of the ac bias the inhomogeneity of the junction
becomes significant, and the ideal model accounted by Eq.
(\ref{sG1}) is no longer valid. Numerical simulations of Eq.
(\ref{sG1}) with boundary conditions (\ref{bc}) have been
performed with the 4th order Runge-Kutta method. The average
voltage drop has been computed and plotted in Figs.
~\ref{phase-shift} (a,b) with the solid line. The following
experimental values of the systems parameters have been taken:
$\alpha=0.05$, $\Omega=0.01$, $\tilde \gamma_1=\tilde
\gamma_2=0.1$ for the case of Fig. \ref{phase-shift}(a) and
$\Omega=0.02$, $\tilde \gamma_1=\tilde \gamma_2=0.135$ for the
case of Fig. \ref{phase-shift}(b).

The results of the experimental measurements and
 numerical simulations are in good correspondence with
the results of the first order (point-particle approximation)
soliton perturbation theory\cite{pert}. In this approximation the
perturbation is assumed to be small, so that it does not change
the shape of the soliton and only the soliton parameters vary in
time. Also, this approximation works well in the adiabatic limit,
when $\Omega \ll \alpha$. Using this approach, the mean fluxon
velocity (in the absence of dc bias) can be
computed\cite{Saler-Zol-2002} as follows:
\begin{equation}
\langle v \rangle \sim \tilde \gamma_1^2\tilde \gamma_2
\sin(\theta+ \theta_0),~ \tan \theta_0 = \frac{2}{{\alpha \over
\Omega} \left [3 +{\left (\alpha \over \Omega \right )}^2
\right]}. \label{apprx}
\end{equation}
The voltage drop is proportional to the average fluxon velocity.
Thus, the rectified voltage behaves as a $sine$ function of the
phase shift. Note that  $\theta_0 \rightarrow 0$ if $\Omega
\rightarrow 0$, and this is in accordance with experimental and
numerical data, where the parameter $\theta_0$ is small.

\begin{figure}[tbp]
\centerline{\includegraphics[width=7.cm,height=6.0cm,clip]{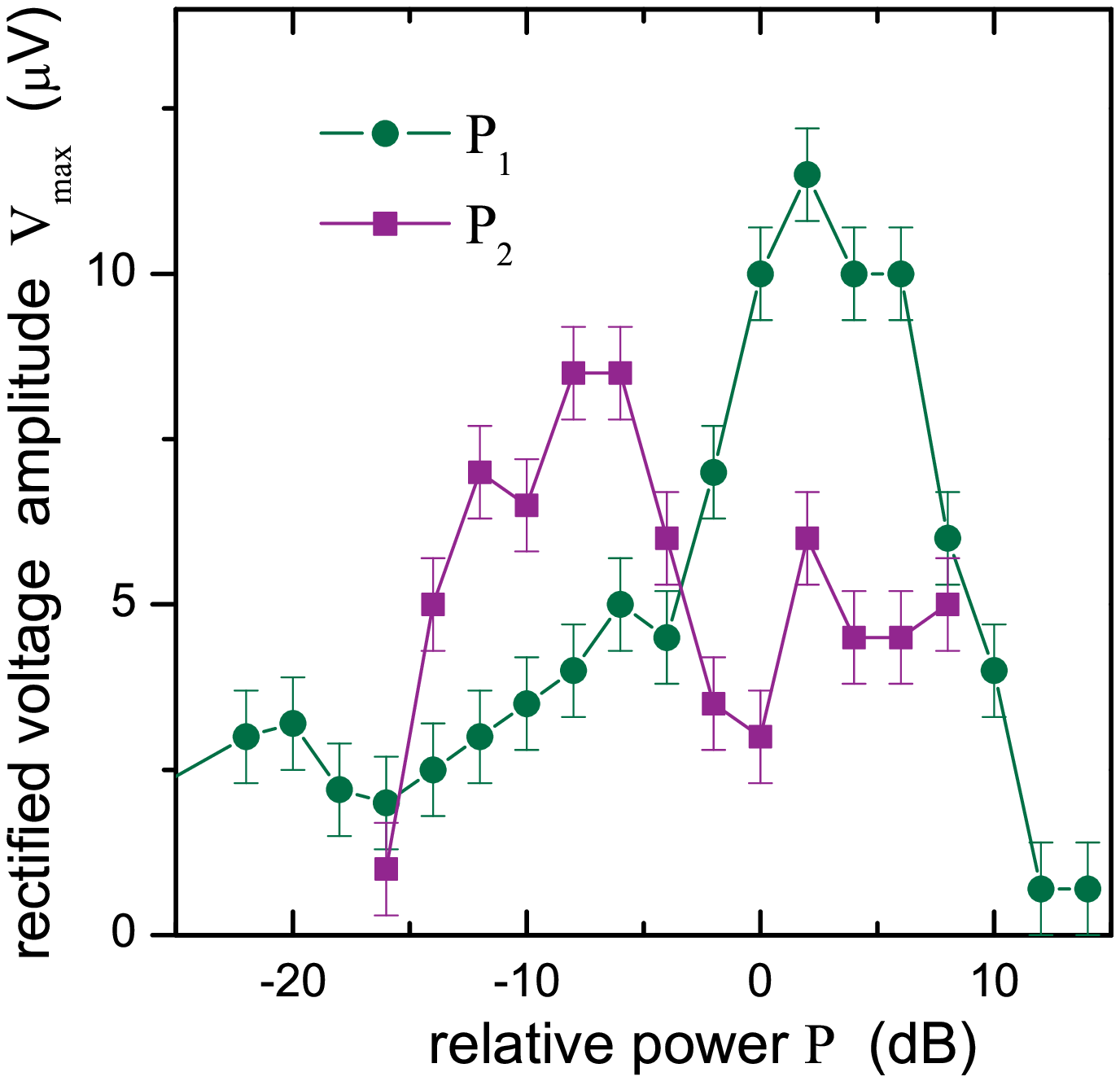}}
\caption{A dependence of the maximum amplitude of the rectified dc
voltage versus power for the two harmonics of ac drive.}
\label{power}
\end{figure}

The maximum amplitude of the rectified voltage is rather large --
it corresponds to the average normalized fluxon velocity of about
0.2. Figure~\ref{power} shows the power dependence of the maximum
amplitude of the rectified dc voltage. For each of the two curves
the power level of another harmonic was kept constant (but not
referenced). One can see that the rectified voltage has a maximum
value at some optimum power level.  The most important system
parameter for the phenomenon is the normal state resistance of the
junction. We find that by decreasing the damping in the system
both the optimal power level and for maximum rectified voltage
increase. The dependence of the bi-harmonic drive on the
phenomenon can be qualitatively undertstood in a simple way. In
the limit of the small rf frequency $f$ (with respect to the
plasma frequency $f_p$) the kink behaves as a point-like particle
and the system reacts to the rf drive adiabatically.
 For any chosen fundamental period of the microwave, the
phase shift between two drives makes the total current large and
positive over short time and then smaller and negative over
somewhat longer time. Since the current amplitude is large enough
to reach the saturation voltage for both current polarities
(Swihart velocity in the case of fluxon), the mean voltage will be
negative in this case. Changing the phase shift simply reverses
the polarities of the currents and voltage. For higher
frequencies, however,  the adiabatic approach does not work and
the phenomenon becomes more complicated. In particular, besides
the point particle contribution characterizing the adiabatic
regime, there is an equally important contribution to the drift
current coming from internal oscillations of the kink (internal
mode mechanism). These oscillations are asymmetric in space and
phase locked to the external driver and, in presence of damping,
they are coupled to the center of mass (translational mode) giving
an extra contribution to the transport \cite{Saler-Zol-2002}. It
would be interesting to separate these two contributions, a task
which is not easy to perform  within our present experimental
setting.

In conclusion, we reported the first experimental observation of
the rectified dc voltage, induced by the ratchet-like motion of a
fluxon in uniform annular Josephson junction driven by bi-harmonic
microwave fields.
We find that the symmetry breaking in the rf field $\tilde
\gamma(t)$ leads to the unidirectional motion of a fluxon, which
is manifested in the non-zero rectified voltage across the
junction at zero bias current. Our results qualitatively confirm
the theoretical predictions made in
Refs.~\onlinecite{Fl-Zol-Fist-2002,Saler-Zol-2002}. The rectified
voltage changes periodically with the phase shift $\theta$. The
authors acknowledge discussions with P. L. Christiansen, M.
Fistul, S. Flach, and M. R. Samuelsen. M. S. acknowledges partial
support from a MURST-PRIN-2003 Initiative, and from the European
grant LOCNET no. HPRN-CT-1999-00163.  Y.Z. wishes to acknowledge
support from the INTAS Young Scientist Fellowship no. 03-55-1799.

%
%


\begin{references}
\bibitem{Hypres} Using our layout, the samples were fabricated
at the foundry of HYPRES Inc., Elmsford, NY 10523.
\end{references}
\end{document}